\documentstyle{article}

\title{Continuous approximation of binomial lattices}

\author{V. Grassi, R. A. Leo, G. Soliani and L. Solombrino \\
     Dipartimento di Fisica dell'Universit\`a,\\
     73100 Lecce, Italy,\\
     and Istituto Nazionale di Fisica Nucleare,\\
     Sezione di Lecce, Italy}

\begin{document}

\maketitle
\begin{abstract}

A systematic analysis of a continuous version of a binomial lattice,
containing a real parameter $\gamma$ and covering the Toda field equation as 
$\gamma\rightarrow\infty$, is carried out in the framework of group theory. 
The symmetry algebra of the equation is derived. Reductions by 
one--dimensional and two--dimensional subalgebras of the symmetry algebra and 
their corresponding subgroups, yield notable field  equations in lower 
dimensions whose solutions allow to find exact solutions to the original 
equation. Some reduced equations turn out to be related to potentials of 
physical interest, such as the Fermi--Pasta--Ulam and the Killingbeck 
potentials, and others. An instanton--like approximate solution is also 
obtained which reproduces the Eguchi--Hanson instanton configuration for 
$\gamma\rightarrow\infty$. Furthermore, the equation under consideration is 
extended to $(n+1)$--dimensions. A spherically symmetric form of this 
equation, studied by means of the symmetry approach, provides conformally 
invariant classes of field equations comprising remarkable special cases. One 
of these $(n=4)$ enables us to establish a connection with the Euclidean
Yang--Mills equations, another appears in the context of Differential 
Geometry in relation to the socalled Yamabe problem. All the properties of 
the reduced equations are shared by the spherically symmetric generalized
field equation. 

\end{abstract}

\newpage

\section{Introduction} 
\setcounter{equation}{0}

The continuous (or long--wave) approximations of chains of particles can 
provide a fruitful theoretical framework to be used as a guide in the study 
of the corresponding original discrete systems. These models are generally 
formulated as nonlinear field equations. When such equations are 
integrable, exact solutions can be found as, for example, solitons, 
vortices and shock--wave.\par
Continuous limits of lattice systems may be considered as phenomenological 
models with a proper identity, and their study could be interesting apart 
from the reduction procedure applied in passing to the continuous 
representation \cite{M}. This feature is shared by different models, among 
which the well--known continuous form
\begin{equation}
u_{xx}+u_{yy}-k(e^u)_{zz}=0
\label{I.1}
\end{equation}
of the two--dimensional discrete Toda equation, where $u=u(x,\,y,\,z)$, 
plays a basic role \cite{S}. (Subscripts denote partial derivatives and 
$k=\pm 1$).\par
Indeed, Eq. (\ref{I.1}) appears in a variety of physical areas, running from 
the theory of Hamiltonian systems to general relativity \cite{FS}, 
\cite{FP}. In the latter context, Eq. (\ref{I.1}) occurs in the theory of 
self--dual Einstein spaces, where is rechristened heavenly equation 
\cite{FP}, \cite{P}. This emerges as a limit case from the Toda molecule 
equation \cite{P}, which is exactly integrable \cite{K}.\par
Another nonlinear field equation associated with a nonlinear lattice which 
deserves to be investigated is
\begin{equation}
\Delta\equiv u_{xx}+u_{yy}-k\left[\left(1+{u\over \gamma}\right)^{\gamma-1}
\right]_{zz}=0,
\label{I.2}
\end{equation}
where $\gamma$ is a (real) parameter such that $\gamma\neq 0,\,1$.\par
Really, after suitable rescalings, Eq. (\ref{I.2}) can be interpreted as 
the continuous limit of a uniform two--dimensional nonlinear lattice of $N$ 
particles interacting through the nearest--neighbor potential \cite{LLS}
\begin{equation}
\phi(r_n)={{a_n}\over{b_n}}\left[\left(1+{{b_n r_n}\over\gamma}
\right)^\gamma-(1+b_n r_n)\right],
\label{I.3}
\end{equation}
where $a_n$ and $b_n$ are constants of the $n$--th nonlinear spring, 
$r_n=y_n-y_{n+1}$, and $y_n$ is the displacement of the $n$--particle from 
its equilibrium position.\par
We shall call "binomial lattice" the chain described by the potential 
(\ref{I.3}).\par
We notice that the function (\ref{I.3}) covers the Toda potential 
($\gamma\rightarrow\infty$) \cite{T}, the harmonic potential ($\gamma=2$), 
and the potential used by Fermi, Pasta and Ulam in their computer 
experiment in the early's 1950 ($\gamma=3$) \cite{FPU}.\par
As one expects, for $\gamma\rightarrow\infty$ Eq. (\ref{I.2}) becomes Eq. 
(\ref{I.1}).\par
The purpose of this paper is both to find the symmetry structure and exact 
solutions of Eq. (\ref{I.2}). An effective method to get insight into the 
symmetry properties of Eq. (\ref{I.2}) and, consequently, to construct 
explicit configurations, is based on the reduction approach \cite{O}, which 
exploits group--theoretical techniques. Applying this procedure, we obtain 
the symmetry algebra of Eq. (\ref{I.2}). The generator of the corresponding 
symmetry group has been determined in part with the help of a computer, by 
means of the symbolic language MAPLE V Release 4 \cite{CDF}. Our result is 
that Eq. (\ref{I.2}) admits a {\it{finite}}--dimensional Lie group of 
symmetries, i.e. a local group $G$ of transformations acting on the 
independent variables $(x,\,y,\,z)$ and the dependent variable $u$ with the 
property that whenever $u(x,\,y,\,z)$ is a solution of Eq. (\ref{I.2}), then 
$u'=(g\circ u)(x',\,y',\,z')$ is also a solution for any $g\in G$.\par
We remind the reader that in contrast with what happens for Eq. (\ref{I.2}), 
the Toda equation (\ref{I.1}), handled within the group theory \cite{Alf}, 
allows an {\it{infinite}}--dimensional symmetry algebra, a realization of 
which is given by generators obeying a Virasoro algebra without a central 
charge (Witt algebra). Furthermore, certain reduced equations give rise to 
instanton and meron--like solutions endowed with integer and fractional 
topological numbers, respectively.\par
Thus, the comparative analysis of the properties of Eqs. (\ref{I.1}) and 
(\ref{I.2}) is of particular relevance, keeping in mind also the fact that 
both the equations come from the continuous approximation of physically 
significant lattice systems.\par
Another important characteristic common to Eqs. (\ref{I.1}) and (\ref{I.2}) 
is that they are related to nonlinear systems of the hydrodynamical type 
\cite{FS}. Although interesting, here this aspect will not be treated.\par
The main results achieved in this article are presented in the Sections in 
which the paper is organized.\par
In Sec. 2 we find the symmetry algebra of Eq. (\ref{I.2}). It turns out that 
the related independent infinitesimal operators obey a closed set of 
commutation rules. Starting from special linear combinations of these 
operators, in Sec. 3 the corresponding symmetry group transformations are 
derived. These provide reduced differential equations leading to exact 
solutions to Eq. (\ref{I.2}). Notable cases arise for $\gamma=3,\,-2,\,
{5\over 2},\,{5\over 3}$. This choice of values is motivated by the fact 
that, in correspondence, the function (\ref{I.3}) coincides with potentials 
of physical interest, such as the Fermi--Pasta--Ulam $(\gamma=3)$ \cite{FPU} 
and a potential involved in the Thomas--Fermi model of an atom $(\gamma=5/2)$ 
(\cite{F}, p. 116). We point out that for $\gamma=3$, a remarkable 
representation of the inverse Weierstrass function is given in terms of the 
Gauss hypergeometric function. In Sec. 4, we obtain an approximate 
instanton--like configuration, holding for large values of the parameter 
$\gamma$, which reproduces the Eguchi--Hanson instanton solution of Eq. 
(\ref{I.1}) in the limit $\gamma\rightarrow\infty$ \cite{EH}.\par
In Sec. 5 we extend Eq. (\ref{I.2}) to the $(n+1)$--dimensional case. The 
exploration of the symmetry properties of this more general nonlinear field 
equation, that is Eq. (\ref{V.1}), has been suggested by the purpose of 
ascertaining whether a link might exist between Eq. (\ref{V.1}) and certain 
problems inherent in Yang--Mills theory and Differential Geometry \cite{Y},
\cite{G}, \cite{tH}.\par
Precisely, we show that applying the reduction technique to the spherically 
symmetric version of Eq. (\ref{V.1}), a connection can be established 
between solutions of Eq. (\ref{V.1}) and solutions of a reduced nonlinear 
ordinary differential equation, i.e. Eq. (\ref{V.13}), which is invariant under 
conformal transformations. Equation (\ref{V.13}) comprises interesting cases, 
which come from $n=1,\,3,\,4,\,6$ and 10 respectively. For instance, for 
$n=10$, Eq. (\ref{V.13}) can be interpreted as an extended (elliptic) form of 
the equation governing the Thomas--Fermi model of an atom, while for $n=4$, 
Eq. (\ref{V.13}) turns out to be related to Euclidean Yang--Mills equations 
via 't Hooft's ansatz \cite{tH}. The properties of Eq. (\ref{V.13}) can be 
reflected on those of Eq. (\ref{V.1}). This is illustrated in Sec. 6.\par
Finally, Sec. 7 contains some concluding remarks, while in Appendixes A, B, 
C and D details of the calculation are reported.

\section{Group analysis}
\setcounter{equation}{0}

To the aim of looking for the symmetry algebra of Eq. (\ref{I.2}), we shall 
apply the standard procedure described in \cite{O}. In doing so, let us 
introduce the vector field 
\begin{equation}
V=\xi\partial_x+\eta\partial_y+\zeta\partial_z+\phi\partial_u,
\label{II.1}
\end{equation}
where $\xi$, $\eta$, $\zeta$ and $\phi$ are functions of $x$, $y$, $z$, 
$u$, and $\partial_x={\partial\over{\partial x}}$, and so on.\par
A local group of transformations is a symmetry group for Eq. (\ref{I.2}), if 
and only if 
\begin{equation}
pr^{(2)}V[\Delta]=0,
\label{II.2}
\end{equation}
whenever $\Delta=0$ for every generator $V$ of $G$, where $pr^{(2)}V$ is 
the second prolongation of $V$ \cite{O}.\par
Equation (\ref{II.2}) give rise to a set of constraints in the form of partial 
differential equations which enable us to determine the coefficients $\xi$, 
$\eta$, $\zeta$ and $\phi$. This has been carried out in part using a 
computer, by means of the symbolic language of \cite{CDF}. However, to 
facilitate the understanding of the method to non specialist readers, in 
Appendix A we write $pr^{(2)}V$ explicily.\par
A general element of the symmetry algebra of Eq. (\ref{I.2}) is
\begin{equation}
V=(c_1 x+c_2 y+c_3)\partial_x+(-c_2 x+c_1 y+c_4)\partial_y+
(c_5+c_6z)\partial_z+2{{\gamma+u}\over{\gamma-2}}(c_6-c_1)\partial_u,
\label{II.3}
\end{equation}
where $c_1$, $c_2$, ..., $c_6$ are arbitrary constants. The expression 
(\ref{II.3}) holds for any allowed value of $\gamma$ $(\gamma\neq 0,\,1)$, 
except for $\gamma={2\over 3},\,2$. We shall come back to these cases 
later.\par
From (\ref{II.3}) we obtain the following independent generators of the 
symmetries of Eq. (\ref{I.2}):
\begin{eqnarray}
&V_1=x\partial_x+y\partial_y-2{{\gamma+u}\over{\gamma-2}}\partial_u,
 \label{II.4a}\\
&V_2=y\partial_x-x\partial_y,\label{II.4b}\\
&V_3=\partial_x,\label{II.4c}\\
&V_4=\partial_y,\label{II.4d}\\
&V_5=\partial_z,\label{II.4e}\\
&V_6=z\partial_z+2{{\gamma+u}\over{\gamma-2}}\partial_u.\label{II.4f}
\end{eqnarray}\par
These operators satisfy the commutation relations
\begin{eqnarray}
&[V_1,\,V_2]=[V_1,\,V_5]=[V_1,\,V_6]=[V_2,\,V_5]=[V_2,\,V_6]=\nonumber\\
&=[V_3,\,V_4]=[V_3,\,V_5]=[V_3,\,V_6]=[V_4,\,V_5]=[V_4,\,V_6]=0,
 \label{II.5a}\\
&[V_1,\,V_3]=-V_3,\label{II.5b}\\
&[V_1,\,V_4]=-V_4,\label{II.5c}\\
&[V_2,\,V_3]=V_4,\label{II.5d}\\
&[V_5,\,V_6]=V_5.\label{II.5e}
\end{eqnarray}
We deduce that the operators (\ref{II.4a})--(\ref{II.4f}) form the basis of a 
six--dimensional solvable Lie algebra ${\cal{L}}$ containing a 
three--dimensional ideal $\{ V_2,\,V_3,\,V_4\}$ which is isomorphic to the 
algebra of $E_2$, the Euclidean group in the plane. The center of 
${\cal{L}}$ is zero, while its derived algebra is $\{ V_2,\,V_3,\,V_4\}$. 
Furthermore, ${\cal{L}}$ admits the Levi decomposition ${\cal{L}}=
{\cal{L}}_1\rhd {\cal{L}}_2$, where ${\cal{L}}_1=\{V_3,\,V_4,\,V_5\}$ and
${\cal{L}}_2=\{V_1,\,V_2,\,V_6\}$. This decomposition is not unique. We can 
easily write other decompositions of this kind). The symbol $\rhd$
stands for the operation of semidirect sum. (To keep the length of the 
paper reasonable, here we shall not explain the algebraic terminology used 
above. Anyway, the reader unfamiliar with the previous mathematical 
concepts, could address, for instance, to \cite{W}).\par
From the elements $V_1$ and $V_6$ we give the generator of the coordinate 
scale transformation, that is $V_0=V_1+V_6=x\partial_x+y\partial_y+
z\partial_z$; $V_3$, $V_4$, $V_5$ generate $x$, $y$ and $z$--translations, 
$V_2$ is a rotation symmetry operator, and $V_1$ and $V_6$ have the meaning 
of generators of two dilations together with a translation of 
$2{\gamma\over{\gamma-2}}$ along the $\mp u$ directions, respectively.\par
As we have already mentioned, the symmetry generator (\ref{II.3}) does not 
include the value $\gamma={2\over 3}$. This is due to the fact that a 
special case emerges, just for $\gamma={2\over 3}$, in solving the 
equations determining the coefficients involved in (\ref{II.1}). In this 
case, to the generators (\ref{II.4a})--(\ref{II.4f}) of the symmetries of Eq. 
(\ref{I.2}), one needs to add another operator, namely
\begin{equation}
V_7=z^2\partial_z-z(2+3u)\partial_u,
\label{II.6}
\end{equation}
which commutes with $V_1$, $V_2$, $V_3$, $V_4$, while
\begin{equation}
[V_6,\,V_7]=V_7,\;\;\;[V_5,\,V_7]=2V_6.
\label{II.7}
\end{equation}\par
A second special case to be dealt with separately is $\gamma=2$, in 
correspondence of which Eq. (\ref{I.2}) becomes the linear wave equation
\begin{equation}
u_{xx}+u_{yy}-{k\over 2}u_{zz}=0.
\label{II.8}
\end{equation}
This equation is the continuous approximation of the lattice system arising 
from the harmonic potential $\phi(r_n)={1\over 4}a_n b_n r_n^2$ (see 
(\ref{I.3})). The symmetry algebra related to Eq. (\ref{II.8}) is 
represented by the vector fields
\begin{eqnarray}
&G_1=x(z\partial_z+y\partial_y)+{1\over 2}(x^2-y^2+2kz^2)\partial_x-
  {1\over 2}xu\partial_u,\label{II.9a}\\
&G_2=x\partial_x+y\partial_y+z\partial_z,\label{II.9b}\\
&G_3=x\partial_z+2kz\partial_x,\label{II.9c}\\
&G_4=-x\partial_y+y\partial_x,\label{II.9d}\\
&G_5=\partial_y,\label{II.9e}\\
&G_6=\partial_z,\label{II.9f}\\
&G_7=\partial_x,\label{II.9g}\\
&G_8=u\partial_u,\label{II.9h}\\
&G_9=f(x,\,y,\,z)\partial_u,\label{II.9i}
\end{eqnarray}
where $f$ is an arbitrary solution of the wave equation $f_{xx}+f_{yy}-
2kf_{zz}=0$. The commutation rules fulfilled by $G_1$,...,$G_9$ can be 
easily found.\par
For brevity, the case $\gamma={2\over 3}$ and $\gamma=2$ will not be 
further discussed. We remind only the reader interested in going deep into 
the case $\gamma=2$, that an exhaustive analysis of an equation of the type 
(\ref{II.8}) is performed in \cite{O}.

\section{Exact solutions from reduced equations}
\setcounter{equation}{0}

The technique of symmetry reduction of a field equation amounts 
essentially to finding the invariants (symmetry variables) of a given 
subgroup of the symmetry group allowed by the equation under consideration 
\cite{O}. This method is usually preceded by the classification of the 
subalgebras of the symmetry algebra. The classification scheme is based on 
the adjoint action of the symmetry group \cite{W}. However, because of the 
finite--dimensionality of the symmetry algebra (\ref{II.5a})--(\ref{II.5e}), 
we shall adopt a more heuristic procedure. In other words, since for 
practical purposes generally one confines oneself to handle only 
low--dimensional symmetry subalgebras, which can be recognized in our case 
directly by inspection, we do not need to employ the adjoint subgroup 
classification method.\par
For each symmetry group generator $V$, we can obtain a basis set of 
invariants $I(x,\,y,\,z,\,u)$ by solving the first order partial 
differential equation
\begin{equation}
VI(x,\,y,\,z,\,u)=0.
\label{III.1}
\end{equation}\par
In the following, we shall build up the reduction of Eq. (\ref{I.2}) by the 
one--dimensional subalgebras $\{V_1\}$, $\{V_2\}$, $\{V_6\}$, and by the 
two--dimensional subalgebras $\{V_1,\,V_2\}$, $\{V_1,\,V_5\}$, 
$\{V_2,\,V_6\}$.

\subsection{Reductions by two--dimensional subalgebras}

The reduced equations associated with the two--dimensional subalgebras $i)$ 
$\{V_1,\,V_2\}$, $ii)$ $\{V_1,\,V_5\}$ and $iii)$ $\{V_2,\,V_6\}$, turn out 
to be nonlinear ordinary differential equations which can be solved to give 
exact solutions to Eq. (\ref{I.2}).\par
{\it{Case i)}}\par
A basis of invariants of the subgroup of the subalgebra $\{V_1,\,V_2\}$ is 
furnished by the system of equations $V_1I=0$, $V_2I=0$, which have to be 
satisfied simultaneously. We obtain
\begin{equation}I_1=z,\,\,\,I_2\equiv A(z)=(x^2+y^2)^{1/2}
\left( 1+{u\over\gamma}\right)^{{\gamma-2}\over 2},
\label{III.1.1}
\end{equation}
from which the reduced equation
\begin{equation}
AA''-{2\over{2\gamma}}{A'}^2-{{2k\gamma}\over{(\gamma-1)(\gamma-2)}}=0,
\label{III.1.2}
\end{equation}
with $A'={{dA}\over{dz}}$, is derived.\par
Putting
\begin{equation}
A=\theta^{{\gamma-2}\over \gamma}
\label{III.1.3}
\end{equation}
into Eq. (\ref{III.1.2}), we get
\begin{equation}
\theta''={{2k}\over{\gamma-1}}\left( {\gamma\over{2-\gamma}}\right)^2
\theta^{{4-\gamma}\over\gamma},
\label{III.1.4}
\end{equation}
yielding
\begin{equation}
{\theta'}^2={{k\gamma^3}\over{(\gamma-1)(\gamma-2)^2}}
\theta^{4\over \gamma}+c,
\label{III.1.5}
\end{equation}
where $c$ is a constant of integration.\par
Equation (\ref{III.1.5}) leads to the relation
\begin{equation}
\int_0^\theta\left(1+b{\theta'}^{4\over \gamma}\right)^{-{1\over 2}}d\theta'
=\pm\sqrt{c}(z-z_0),
\label{III.1.6}
\end{equation}
$z_0$ being an arbitrary constant, and
\begin{equation}
b={k\over c}{{\gamma^3}\over{(\gamma-1)(\gamma-2)^2}}.
\label{III.1.7}
\end{equation}\par
The integral at the left--hand side of (\ref{III.1.6}) can be calculated 
resorting to the formula (see \cite{GR}, p. 284)
\begin{equation}
\int_0^X{{{X'}^{\mu-1}dX'}\over{(1+bX')^\nu}}={{X^\mu}\over\mu}
{_{2}F_{1}}(\nu,\,\mu;\,\mu+1;\,-bX),
\label{III.1.8}
\end{equation}
where ${\rm{arg}}(1+bX)<\pi$, $Re\mu>0$ and ${_{2}F_{1}}$ denotes the Gauss 
hypergeometric function. Indeed, setting in (\ref{III.1.6}) 
$X=\theta^{4\over\gamma}$ and using (\ref{III.1.8}) with 
$\mu={\gamma\over 4}$ and $\nu={1\over 2}$, we find
\begin{equation}
\theta{_{2}F_{1}}\left({1\over 2},\,{\gamma\over 4};\,{\gamma\over 4}+1;\,
-b\theta^{4\over\gamma}\right)=\pm\sqrt{c}(z-z_0),
\label{III.1.9}
\end{equation}
where $\theta=\theta(\pm\sqrt{c}(z-z_0))$ is explicitly known whenever Eq. 
(\ref{III.1.9}) is invertible. When this occurs, Eq. (\ref{III.1.1}) provides 
an exact solution to Eq. (\ref{I.2}). For example, for $\gamma=4$, $k=-1$ 
and $z_0=-{3\over 8}\sqrt{c}$, we have 
\begin{equation}
{_{2}F_{1}}\left({1\over 2},\,1;\,2;\,-b\theta\right)=
{{2(-1+\sqrt{1+b\theta})}\over{b\theta}},
\label{III.1.10}
\end{equation}
with $b=-{{16}\over{3c}}$.\par
Then, combining together Eqs. (\ref{III.1.9}) and (\ref{III.1.1}), we obtain 
the solution
\begin{equation}
u=4\left[{{3c}\over{16}}{1\over{x^2+y^2}}\left(1-{{64}\over{9c}}z^2\right)
\right]^{1\over 2}-4
\label{III.1.11}
\end{equation}
to the equation
\begin{equation}
u_{xx}+u_{yy}+\left[\left(1+{u\over 4}\right)^3\right]_{zz}=0.
\label{III.1.12}
\end{equation}
{\it{Case ii)}}\par
The subgroup of the subalgebra $\{V_1,\,V_5\}$ admits the basis of 
invariants
\begin{equation}
I_1\equiv\eta={x\over y},\;\;\;I_2\equiv B(\eta)={1\over z}\left( 1+
{u\over\gamma}\right)^{{2-\gamma}\over 2}.
\label{III.1.13}
\end{equation}
Then, the reduced equation of Eq. (\ref{I.2}) coming from the symmetry 
variables $\eta$ and $B(\eta)$ takes the form
\begin{equation}
\eta^2(\eta^2+1){{B''}\over B}+2\eta\left(\eta^2+{2\over{2-\gamma}}\right)
{{B'}\over B}+{\gamma\over{2-\gamma}}\left[\eta^2(\eta^2+1)\left(
{{B'}\over B}\right)^2+1\right]=0,
\label{III.1.14}
\end{equation}
with $B'={{dB}\over{d\eta}}$.\par
By introducing the change of variable
\begin{equation}
{{B'}\over B}={{2-\gamma}\over 2}{{Q'}\over Q},
\label{III.1.15}
\end{equation}
Eq. (\ref{III.1.14}) can be cast into the linear differential equation
\begin{equation}
\eta^2(\eta^2+1)Q''+2\eta\left(\eta^2+{2\over{2-\gamma}}\right)Q'+
{{2\gamma}\over{(2-\gamma)^2}}Q=0.
\label{III.1.16}
\end{equation}
Resorting to the MATHEMATICA symbolic language, we find the 
general solution
\begin{equation}
Q(\eta)=\left({{1+\eta^2}\over{\eta^2}}\right)^{1\over{2-\gamma}}\left[
c_1\sin\left({2\over{2-\gamma}}{\rm{arctang}}\eta\right)+c_2\cos\left(
{2\over{2-\gamma}}{\rm{arctang}}\eta\right)\right]
\label{III.1.17}
\end{equation}
of Eq. (\ref{III.1.16}), where $c_1$ and $c_2$ are arbitrary constants.\par
Thus, by virtue of (\ref{III.1.17}) and (\ref{III.1.15}), an exact solution of 
Eq. (\ref{I.2}) is determined, namely
\begin{equation}
u=\gamma\left( z^{2\over{2-\gamma}}Q-1\right).
\label{III.1.18}
\end{equation}
{\it{Case iii)}}\par
A set of basis invariants for the subgroup of the subalgebra 
$\{V_2,\,V_6\}$ is
\begin{equation}
I_1\equiv\tau=(x^2+y^2)^{1\over 2},\;\;\;I_2\equiv G(\tau)=z\left(1+
{u\over\gamma}\right)^{{2-\gamma}\over 2}.
\label{III.1.19}
\end{equation}
These give rise to the reduced equation
\begin{equation}
{\gamma\over{2-\gamma}}{G'}^2+{{GG'}\over\tau}+GG''=k
{{\gamma-1}\over{2-\gamma}}
\label{III.1.20}
\end{equation}
of Eq. (\ref{I.2}), with $G'={{dG}\over{d\tau}}$.\par
Equation (\ref{III.1.20}) can be written as
\begin{equation}
\theta''+{1\over\tau}\theta'=2k{{\gamma-1}\over{(2-\gamma)^2}}
\theta^{\gamma-1}
\label{III.1.21}
\end{equation}
via the transformation
\begin{equation}
G=\theta^{{2-\gamma}\over 2}.
\label{III.1.22}
\end{equation}
Thus, a particular solution of Eq. (\ref{III.1.21}) is
\begin{equation}
\theta=\left({{2k}\over{\gamma-1}}{1\over{\tau^2}}\right)^
{1\over{\gamma-2}},
\label{III.1.23}
\end{equation}
which yields (see (\ref{III.1.19}) and (\ref{III.1.23}))
\begin{equation}
u=\gamma\left[\left({{2k}\over{\gamma-1}}{\tau\over z}\right)^
{2\over{2-\gamma}}-1\right].
\label{III.1.24}
\end{equation}

\subsection{Reductions by one--dimensional subalgebras}

In opposition to what happens in the case of two--dimensional (symmetry) 
subalgebras, the one--dimensional subalgebras provide reduced equations of 
(\ref{I.2}) given by partial differential equations in two independent 
variables. Hence, in order to arrive at reduced ordinary differential 
equations, we have to apply again the reduction procedure. This emerges in 
dealing with the one--dimensional subalgebras $i)$ $\{V_1\}$, $ii)$ 
$\{V_2\}$ and $iii)$ $\{V_6\}$.\par\noindent
{\it{Case i)}}\par
Associated with the symmetry vector field $V_1$, the basis of invariants 
\begin{equation}
I_1=x,\;\;\;I_2\equiv\eta={x\over y},\;\;\;
I_3\equiv A(\eta,\,z)={1\over x}\left(1+{u\over\gamma}\right)^
{{2-\gamma}\over 2},
\label{III.2.1}
\end{equation}
can be constructed. Then, substitution from the variables (\ref{III.2.1}) 
into Eq. (\ref{I.2}) yields the reduced equation
\begin{eqnarray}
&\eta^2(\eta^2+1)AA''+2\eta(\eta^2+a)AA'+(a-1)\eta^2(\eta^2+1){A'}^2+(a-1)
 A^2=\nonumber\\
&{k\over 2}{{a-2}\over{a-1}}\left[(a-3)\left({{A_x}\over A}\right)^2+
 {{A_{xx}}\over A}\right],\label{III.2.2}
\end{eqnarray}
where $A'={{\partial A}\over{\partial\eta}}$, 
$A_x={{\partial A}\over{\partial x}}$ and $a={2\over{2-\gamma}}$.\par
The symmetry algebra of Eq. (\ref{III.2.2}) is expressed by the generators
\begin{eqnarray}
&T_1=z\partial_z-u\partial_u,\label{III.2.3a}\\
&T_2=(\eta^2+1)\partial_\eta-{u\over\eta}\partial_u,\label{III.2.3b}\\
&T_3=\partial_z.\label{III.2.3c}
\end{eqnarray}
The reduction by the vector field $T_1$ leads to the invariants
\begin{equation}
I_1=\eta,\;\;\;I_2\equiv U(\eta)=zA(\eta,\,z),
\label{III.2.4}
\end{equation}
which once replaced into Eq. (\ref{III.2.2}) provide the ordinary 
differential equation
\begin{equation}
\eta^2(\eta^2+1)UU''+2\eta(\eta^2+a)UU'+(a-1)\eta^2(\eta^2+1){U'}^2+(a-1)U^2
={k\over 2}(a-2),
\label{III.2.5}
\end{equation}
with $U'={{dU}\over{d\eta}}$. This equation can be simplified by putting
\begin{equation}
U=\theta^{1\over a}.
\label{III.2.6}
\end{equation}
Indeed, we have
\begin{equation}
\eta^2(\eta^2+1)\theta''+2\eta(\eta^2+a)\theta'+a(a-1)\theta={k\over 2}
a(a-2)\theta^{{a-2}\over a},
\label{III.2.7}
\end{equation}
where the following
\begin{equation}
\theta=\left({k\over 2}{{a-2}\over{a-1}}\right)^{a\over 2}{1\over{\eta^a}}
\label{III.2.8}
\end{equation}
is a special solution. Now, taking account of (\ref{III.2.8}), 
(\ref{III.2.6}), (\ref{III.2.4}) and (\ref{III.2.1}), we find
\begin{equation}
u=\gamma\left[\left({{\gamma-1}\over\gamma}{{y^2}\over{z^2}}\right)^
{1\over{2-\gamma}}-1\right].
\label{III.2.9}
\end{equation}
Furthermore, we observe that (\ref{III.2.5}) allows the constant solution
$U=\sqrt{{k\over 2}{{a-2}\over{a-1}}}$, to which the solution (symmetric of 
(\ref{III.2.9}) with respect to the exchange $y\rightarrow x$):
\begin{equation}
u=\gamma\left[\left({{\gamma-1}\over\gamma}{{x^2}\over{z^2}}\right)^
{1\over{2-\gamma}}-1\right]
\label{III.2.10}
\end{equation}
to Eq. (\ref{I.2}) corresponds.\par
Another interesting reduced equation can be derived from the symmetry 
generator $T_2$. The related invariants are
\begin{equation}
I_1=z,\;\;\;I_2\equiv W(z)={{A(x,\,\eta)\eta}\over{\sqrt{\eta^2+1}}}.
\label{III.2.11}
\end{equation}
The introduction of (\ref{III.2.11}) into Eq. (\ref{III.2.2}) gives
\begin{equation}
W''+(a-3)W'=2k{{a(a-1)}\over{a-2}}W^3,
\label{III.2.12}
\end{equation}
with $W'={{dW}\over{dz}}$.\par
For $a=3$ $\left(\gamma={4\over 3}\right)$, Eq. (\ref{III.2.12}) becomes
\begin{equation}
W''=12kW^3,
\label{III.2.13}
\end{equation}
which affords the solution
\begin{equation}
W={1\over{\sqrt 6}}{1\over{z-z_0}}
\label{III.2.14}
\end{equation}
for $k=-1$, $z_0$ being an arbitrary constant.\par
Another solution of Eq. (\ref{III.2.13}) can be obtained carrying out a first 
integration, which furnishes
\begin{equation}
{W'}^2=6kW^4+const.
\label{III.2.15}
\end{equation}
A second integration provides a solution expressed in terms of elliptic 
functions (\cite{Ka}, p.543).\par
In general, by setting $W=\xi^{3-a}\sigma(\xi)$, $\xi=e^z$, Eq. 
(\ref{III.2.12}) can be written as
\begin{equation}
\xi\sigma_{\xi\xi}+(4-a)\sigma_\xi-2k{{a(a-1)}\over{a-2}}
\xi^{-2a+5}\sigma^3=0,
\label{III.2.16}
\end{equation}
that is an equation belonging to the class
\begin{equation}
\xi\sigma_{\xi\xi}+a_1\sigma_\xi+a_2\xi^\nu\sigma^n=0
\label{III.2.17}
\end{equation}
($a_1$, $a_2$ are constants) studied by Flower and by other authors, the 
references of them are quoted in (\cite{Ka}, p. 560).\par\noindent
{\it{Case ii)}}\par
The symmetry subalgebra $\{V_2\}$ gives rise to the reduced equation
\begin{equation}
u_{\tau\tau}+{1\over\tau}u_\tau=k\left[\left(1+{u\over\gamma}\right)^
{\gamma-1}\right]_{zz},
\label{III.2.18}
\end{equation}
where the basis of invariants
\begin{equation}
I_1\equiv\tau=\sqrt{x^2+y^2},\;\;\;I_2\equiv u(\tau,\,z)
\label{III.2.19}
\end{equation}
has been used.\par
A group analysis of Eq. (\ref{III.2.18}) provides the vector fields
\begin{eqnarray}
&S_1=z\partial_z+\tau\partial_\tau\label{III.2.20a}\\
&S_2={{\gamma-2}\over 2}z\partial_z+(\gamma+u)\partial_u,\label{III.2.20b}\\
&S_3=\partial_z.\label{III.2.20c}
\end{eqnarray}
We note that for $\gamma={2\over 3}$, in addition to $S_1$, $S_2$, $S_3$ a 
further generator exists, i.e.
\begin{equation}
S_4=\left({2\over 3}+u\right)z\partial_u-{1\over 3}z^2\partial_z.
\label{III.2.21}
\end{equation}
The reduced equation coming from (\ref{III.2.20a}) reads
\begin{equation}
\rho^2V''+\rho V'={k\over\gamma}{{d^2}\over{d\rho^2}}V^{\gamma-1},
\label{III.2.22}
\end{equation}
where
\begin{equation}
V=1+{u\over\gamma},\;\;\;\rho={z\over\tau},\;\;\;u=u(\rho).
\label{III.2.23}
\end{equation}
A particular solution of Eq. (\ref{III.2.22}) is
\begin{equation}
V=\left({{2k}\over{\gamma-1}}\right)^{1\over{\gamma-2}}
\rho^{2\over{\gamma-2}},
\label{III.2.24}
\end{equation}
which yields
\begin{equation}
u=\gamma\left[\left( {{2k}\over{\gamma-1}} \right)^{1\over{\gamma-2}}
\left( {z\over\tau} \right)^{2\over{\gamma-2}}-1\right]
\label{III.2.25}
\end{equation}
from (\ref{III.2.23}), with $\tau=\sqrt{x^2+y^2}$.\par
On the other hand, the generator (\ref{III.2.20b}) leads to the reduced 
equation
\begin{equation}
W''+{1\over\tau}W'=\lambda W^{\gamma-1},
\label{III.2.26}
\end{equation}
where $w(\tau)={{\gamma+u}\over{z^{2\over{\gamma-2}}}}$, 
$\tau=\sqrt{x^2+y^2}$, $W'={{dW}\over{d\tau}}$ and
$\lambda={{2k}\over{\gamma^{\gamma-2}}}{{\gamma-1}\over{(\gamma-2)^2}}$.
\par
A solution of Eq. (\ref{III.2.26}) is
\begin{equation}
W=W_0\tau^{-{2\over{\gamma-2}}},
\label{III.2.27}
\end{equation}
with $W_0=\gamma\left({{2k}\over{\gamma-1}}\right)^{1\over{\gamma-2}}$. In 
correspondence of (\ref{III.2.27}), the following solution
\begin{equation}
u=W_0\left({{z^2}\over{x^2+y^2}}\right)^{1\over{\gamma-2}}-\gamma
\label{III.2.28}
\end{equation}
to Eq. (\ref{I.2}) is found.\par\noindent
{\it{Case iii)}}\par
By solving the equation $V_6I=0$, we get the set of basis invariants
\begin{equation}
I_1=x,\;\;\;I_2=y,\;\;\;I_3\equiv W(x,\,y)=\ln\left[{1\over{z^2}}\left(
1+{u\over\gamma}\right)^{\gamma-2}\right].
\label{III.2.29}
\end{equation}
Then, from (\ref{I.2}) we have the reduced equation
\begin{equation}
W_{xx}+W_{yy}+{1\over{\gamma-2}}(W_x^2+W_y^2)=
{{2k(\gamma-1)}\over{\gamma-2}}\,e^W.
\label{III.2.30}
\end{equation}\par
The group tecnique can be applied to obtain a further reduction of Eq. 
(\ref{III.2.30}). In doing so, we find the symmetry generator
\begin{equation}
N=(-c_1x+c_2y+c_3)\partial_x+(-c_2x-c_1y+c_4)\partial_y+2c_1\partial_W.
\label{III.2.31}
\end{equation}
From (\ref{III.2.31}) we deduce the independent vector fields
\begin{eqnarray}
&N_1=-x\partial_x-y\partial_y+2\partial_W,\label{III.2.32a}\\
&N_2=y\partial_x-x\partial_y,\label{III.2.32b}\\
&N_3=\partial_x,\label{III.2.32c}\\
&N_4=\partial_y,\label{III.2.32d}
\end{eqnarray}
which fulfill the commutation relations
\begin{eqnarray}
&[N_1,\,N_2]=0,\;\;\;[N_1,\,N_3]=N_3,\;\;\;[N_1,\,N_4]=N_4,\label{III.2.33a}\\
&[N_2,\,N_3]=N_4,\;\;\;[N_2,\,N_4]=-N_3,\;\;\;[N_3,\,N_4]=0.\label{III.2.33b}
\end{eqnarray}
It is noteworthy that Eqs. (\ref{III.2.33a})--(\ref{III.2.33b}) define a Lie 
algebra isomorphic to the algebra $sm(2)$ of the similitude group in 
${\cal{R}}^2$, governed by the rules \cite{BK}: $[X_1,\,X_2]=-X_3$, 
$[X_1,\,X_3]=X_2$, $[X_1,\,X_3]=0$, $[X_2,\,X_4]=X_2$, $[X_3,\,X_4]=X_3$, 
$[X_1,\,X_4]=0$. This can be checked by identifying $N_1$, $N_2$, $N_3$, 
$N_4$ with $-X_4$, $-X_1$, $X_2$, $X_3$, respectively.\par
Now let us examine the generator $N_1$, whose a set of symmetry variables is
\begin{equation}
I_1\equiv t={y\over x},\;\;\;I_2\equiv\sigma=2\ln x+W(x,\,y).
\label{III.2.34}
\end{equation}
The reduced equation of (\ref{III.2.30}) emerging from (\ref{III.2.34}) reads
\begin{equation}
(t^2+1)\sigma_{tt}+{1\over{\gamma-2}}(t^2+1)\sigma_t^2+
{{2\gamma}\over{\gamma-2}}t\sigma_t+{{2\gamma}\over{\gamma-2}}=
{{2k(\gamma-1)}\over{\gamma-2}},
\label{III.2.35}
\end{equation}
which is solved by
\begin{equation}
\sigma=\ln\left[{{k\gamma(a^2+b^2)}\over{\gamma-1}}{1\over{(a+bt)^2}}\right],
\label{III.2.36}
\end{equation}
where $a$ and $b$ are constants of integration.\par
Combining together (\ref{III.2.36}), (\ref{III.2.34}) and (\ref{III.2.29}), we 
obtain the class of $\infty^2$--solutions
\begin{equation}
u=\gamma\left\{\left[{{k\gamma(a^2+b^2)}\over{\gamma-1}}
{{z^2}\over{(ax+by)^2}}\right]^{1\over{\gamma-2}}-1\right\}.
\label{III.2.37}
\end{equation}\par
At this stage it is instructive to look for other nontrivial solutions of 
(\ref{I.2}). To this aim, let us set
\begin{equation}
W=(\gamma-2)\ln\psi
\label{III.2.38}
\end{equation}
into Eq. (\ref{III.2.30}). Then, this equation takes the form
\begin{equation}
\psi_{xx}+\psi_{yy}=2k{{\gamma-1}\over{(\gamma-2)^2}}\psi^{\gamma-1}.
\label{III.2.39}
\end{equation}\par
By way of example, we seek a solution of the type $\psi=\psi(\xi)$, where 
$\xi=x-vy$ and $v$ is a (real) constant. In such a manner Eq. (\ref{III.2.39}) 
transforms into the (nonlinear) ordinary differential equation
\begin{equation}
\psi_{\xi\xi}={{2k(\gamma-1)}\over{1+v^2(\gamma-2)^2}}\psi^{\gamma-1},
\label{III.2.40}
\end{equation}
from which
\begin{equation}
\psi_\xi^2=a\psi^\gamma+c,
\label{III.2.41}
\end{equation}
where $c$ is a constant of integration and 
$a={{4k}\over{1+v^2}}{{\gamma-1}\over{\gamma(\gamma-2)^2}}$. For $c=0$ Eq. 
(\ref{III.2.41}) provides
\begin{equation}
\psi=\pm\left[a^{1\over2}\left({{2-\gamma}\over2}\right)\right]^
{2\over{2-\gamma}}(\xi-\xi_0)^{2\over{2-\gamma}},
\label{III.2.42}
\end{equation}
$\xi_0$ being a constant of integration.\par
On the other hand, if $c\neq0$ we obtain
\begin{equation}
\int_0^\psi{{d\psi'}\over{\sqrt{1+b{\psi'}^\gamma}}}=\pm\sqrt{c}
(\xi-\xi_0),
\label{III.2.43}
\end{equation}
with $b={a\over b}$.\par
The integral at the left--hand side of (\ref{III.2.43}) can be evaluated by 
Eq. (\ref{III.1.8}). Indeed, putting in (\ref{III.2.43}) $X=\psi^\gamma$, from 
Eq. (\ref{III.1.8}) we find
\begin{equation}
\gamma\psi{_{2}F_{1}}\left({1\over2},\,{1\over\gamma};\,{{1+\gamma}\over2};
\,-b\psi^\gamma\right)=\pm\sqrt{c}(\xi-\xi_0),
\label{III.2.44}
\end{equation}
with $\mu={1\over\gamma}$ and $\nu={1\over2}$.\par
With the help of (\ref{III.2.29}) and (\ref{III.2.38}), we get the exact 
solution
\begin{equation}
u=\gamma(z^{2\over{\gamma-2}}\psi-1)
\label{III.2.45}
\end{equation}
to Eq. (\ref{I.2}), where $\psi=\psi(x-vy)$ is explicitly known if one can 
invert Eq. (\ref{III.2.44}).\par
Equation (\ref{III.2.44}) covers some cases related to potentials of physical 
interest, arising for a) $\gamma=3$, b) $\gamma=-2$, c) $\gamma={5\over2}$, 
d) $\gamma={5\over 3}$, and e) $\gamma=-1$ (see (\ref{I.3})).\par
The first two choices correspond, respectively, to the Fermi--Pasta--Ulam 
potential \cite{FPU} and to a potential whose nonlinear part, 
$1/\left(1-{u\over2}\right)^2$, finds applications in the treatment of the 
scattering states in conformally invariant Quantum Mechanics \cite{BPS}. In 
cylindric coordinates, case c) leads to an equation involved in the 
Thomas--Fermi model of an atom (\cite{F} p. 116), while in the case d) Eq. 
(\ref{III.2.39}) appears in the theory of the white dwarf stars (\cite{F}, 
p. 113). Finally, for $\gamma=-1$ (case e)) the potential (\ref{I.3}) takes 
the form $\phi\sim(1-u)^{-1}-(1+u)$, which mimics a special case of the 
Killingbeck potential $V=-{A\over r}+Br+Cr^2$ \cite{Ki}. We recall that a 
Coulomb potential perturbed by a second degree polynomial can be used to 
study the ground state properties of a hydrogen atom.\par
For brevity, below we handle in detail only the cases a) and b).\par
\noindent
{\it{Case a)}}\par
For $\gamma=3$, Eq. (\ref{III.2.41}) can be integrated straightforwardly. 
Hence, first we shall evaluate $\psi$ explicitly from (\ref{III.2.41}).\par
Subsequently, we shall compare the expression of $\psi$ determined in such 
a way with that coming from (\ref{III.2.44}). In doing so, let us reshape Eq. 
(\ref{III.2.41}) into
\begin{equation}
\psi_\eta^2=4\psi^3-g_3,
\label{III.2.46}
\end{equation}
after the rescaling $\eta=\sqrt{{2\over{3(1+v^2)}}}\xi$, with
$c=-{2\over{3(1+v^2)}}g_3$ and $k=1$.\par
Equation (\ref{III.2.46}) is a special version of the equation
\begin{equation}
\psi_\eta^2=4\psi^3-g_2\psi-g_3,
\label{III.2.47}
\end{equation}
which is satisfied by the Weierstrass elliptic function 
$\wp(\eta;\,g_2,\,g_3)$, where $g_2$ and $g_3$ are the invariants of 
$\wp$ (\cite{AS} p. 629).\par
For $g_3=0$ we have
\begin{equation}
\psi(\eta)=\wp(\eta;\,0,\,0)={1\over{\eta^2}},
\label{III.2.48}
\end{equation}
that is the first term of the series representation of 
$\wp(\eta;\,g_2,\,g_3)$ (\cite{AS} p. 635). Then, (\ref{III.2.45}) gives
\begin{equation}
u=3\left({{z^2}\over{\eta^2}}-1\right).
\label{III.2.49}
\end{equation}\par
For $g_3\neq 0$, Eq. (\ref{III.2.47}) yields
\begin{equation}
\int_\infty^\psi{{d\psi'}\over{\sqrt{4{\psi'}^3-g_3}}}=\eta,
\label{III.2.50}
\end{equation}
namely
\begin{equation}
\psi=\wp(\eta;\,0,\,g_3).
\label{III.2.51}
\end{equation}
Thus, we infer that
\begin{equation}
u=3\left[z^2\wp(\eta;\,0,\,g_3)-1\right]
\label{III.2.52}
\end{equation}
from (\ref{III.2.45}).\par
Combining (\ref{III.2.51}) together with (\ref{III.2.44}), it can be shown (see 
Appendix B) that the remarkable formula
\begin{equation}
\wp^{-1}=\sigma{_{2}F_{1}}\left({1\over2},\,{1\over3};\,{4\over3};\,
-4\sigma^3\right)-{{4^{1/3}}\over3}B\left({1\over3},\,{1\over6}\right)
\label{III.2.53}
\end{equation}
holds, where $\wp^{-1}$ stands for the inverse Weierstrass function 
$(g_2=0)$, $\sigma=-{\psi\over{g_3^{1/3}}}$ and $B(p,\,q)$ is the beta 
function (Euler's integral of the first kind) (\cite{GR} p. 948). The 
relation (\ref{III.2.53}) estabilishes a link between the Gauss 
hypergeometric function and the inverse Weierstrass function $(g_2=0)$.\par
\noindent
{\it{Case b)}}\par
If $\gamma=-2$, Eq. (\ref{III.2.44}) furnishes
\begin{equation}
-2\psi{_{2}F_{1}}\left({1\over2},\,-{1\over2};\,{1\over2};\,-b\psi^{-2}
\right)=\pm\sqrt{c}(\xi-\xi_0),
\label{III.2.54}
\end{equation}
where $b={{8k}\over{3c(1+v^2)}}$, and
\begin{equation}
{_{2}F_{1}}\left({1\over2},\,-{1\over2};\,{1\over2};\,-b\psi^{-2}\right)=
\sqrt{1+b\psi^{-2}}
\label{III.2.55}
\end{equation}
(see \cite{GR}, p. 1042).\par
Consequently,
\begin{equation}
u=-2\left(\sqrt{{{c(\xi-\xi_0)^2-4b}\over{4z}}}-1\right)
\label{III.2.56}
\end{equation}
from (\ref{III.2.45}).

\section{Instanton--like approximate solution}
\setcounter{equation}{0}

Equation (\ref{III.2.30}) can be exploited as well to give approximate 
solutions to Eq. (\ref{I.2}) of the instanton--like type. To this purpose, 
it is convenient to write Eq. (\ref{III.2.30}) in the form
\begin{equation}
\partial_{\eta}\partial_{\overline{\eta}}W+{1\over{\gamma -2}}W_{\eta}
W_{\overline{\eta}}={k\over 2}{{\gamma-1}\over{\gamma-2}}e^W,
\label{IV.1}
\end{equation}
where $\eta=x+iy$, $\overline{\eta}=x-iy$.\par
For $\gamma\rightarrow \infty$, Eq. (\ref{IV.1}) reproduces the Liouville 
equation
\begin{equation}
\partial_{\eta}\partial_{\overline{\eta}}W={k\over 2}e^W,
\label{IV.2}
\end{equation}
which is conformally invariant, in the sense that if 
$W(\eta,\overline{\eta})$ is a solution of Eq. (\ref{IV.2}), the function
${\widetilde{W}}={\widetilde{W}}(\tilde{\eta},\overline{\tilde{\eta}})$ given 
by
\begin{equation}
W={\widetilde{W}}-\ln f'{\overline{f}}'
\label{IV.3}
\end{equation}
is also a solution, where $\eta=f(\tilde{\eta})$, 
$\overline{\eta}=f(\overline{\tilde{\eta}})$, and 
${f'}=f_{\tilde{\eta}}(\tilde{\eta})$, 
${\overline{f}}'=f_{\overline{\tilde{\eta}}}(\overline{\tilde{\eta}})$.\par
We notice that the Liouville equation emerges as a symmetry reduction of 
the continuous Toda equation (\ref{I.1}) \cite{Alf} corresponding to the 
generator $V_6(\infty)=z\partial_z+2\partial_u$, coming from (\ref{II.4f})
for $\gamma\rightarrow \infty$. Equation (\ref{IV.2}) affords the 
general solution
\begin{equation}
e^W={{4f_{\eta}(\eta)\overline{f}_{\overline{\eta}}(\overline{\eta})}\over
{[1+f(\eta)\overline{f}(\overline{\eta})]^2}}
\label{IV.4}
\end{equation}
for $k=-1$, where $f(\eta)$ denotes an arbitrary holomorphic function. This 
expression can be exploited to get (exact) instanton solutions to Eq. 
(\ref{I.1}) \cite{Alf}, \cite{EH}. Conversely, we are not able to obtain 
exact instanton solutions to Eq. (\ref{IV.1}). However, we can find 
(heuristically) an {\it{approximate}} instanton-like solution, which 
arises from the assumption
\begin{equation}
W=W_{0}+\epsilon\phi,
\label{IV.5}
\end{equation}
where $W_0$ satisfies (\ref{IV.4}), $\epsilon={1\over{\gamma-2}}$ is such 
that $|\epsilon|\ll1$, and $\phi$ is a function of $\eta$, $\overline{\eta}$ to 
be determined.\par
Substitution from (\ref{IV.5}) into (\ref{IV.1}) gives
\begin{equation}
\epsilon^{0} \; : \; W_{0\eta\overline{\eta}}={k\over 2}e^{W_0}
\label{IV.6}
\end{equation}
\begin{equation}
\epsilon \; : \; {\phi}_{\eta\overline{\eta}}+W_{0\eta}W_{0\overline{\eta}}=
{k\over 2}e^{W_0}(1+\phi),
\label{IV.7}
\end{equation}
for large values of $\gamma$. At this point we choose $k=-1$, $f=\eta$, 
$\overline{f}=\overline{\eta}$ in (\ref{IV.4}). Then, $W_0$ becomes the 
instanton configuration
\begin{equation}
e^{W_0}={4\over{(1+\eta \overline{\eta})^2}}.
\label{IV.8}
\end{equation}
With the help of (\ref{IV.8}), Eq. (\ref{IV.7}) can be written as
\begin{equation}
g_{rr}+{1\over r}g_{r}+{1\over{r^2}}g_{\theta\theta}+{8g\over{(1+r^2)^2}}=8
\label{IV.9}
\end{equation}
in polar coordinates ($x=r\cos\theta$, $y=r\sin\theta$), where
\begin{equation}
g(r,\theta)=\phi(r,\theta)+2r^2+1.
\label{IV.10}
\end{equation}\par
We are interested in solutions of (\ref{IV.9}) of the type $g=g(r,0)$. 
Therefore, by introducing the change of variable $r=\left({{1+t}\over{1-t}}
\right)^{1/2}$, Eq. (\ref{IV.9}) takes the form
\begin{equation}
(1-t^2)g_{tt}-2tg_t+2g={8\over{(1-t)^2}}.
\label{IV.11}
\end{equation}
It is noteworthy that, in general, this equation can be solved exactly. In 
fact, its homogeneous part coincides with a special case of the 
hypergeometric equation defining the Legendre functions (\cite{AS}, p. 331). 
Precisely, the homogeneous part of (\ref{IV.11}) allows the independent 
integrals
\begin{equation}
g_1(t)=P_1(t)\equiv t,
\label{IV.12a}
\end{equation}
\begin{equation}
g_2(t)=Q_1(t)\equiv {1\over 2}t\ln{{t+1}\over{t-1}}-1,
\label{IV.12b}
\end{equation}
where $P_1$ and $Q_1$ are the Legendre polynomial and the Legendre function 
of the second kind, respectively, corresponding to $n=1$.\par
In terms of $r^2$, we have
\begin{equation}
g_1={{r^2-1}\over{r^2+1}},
\label{IV.13a}
\end{equation}
\begin{equation}
g_2={1\over 2}{{r^2-1}\over{r^2+1}}\ln {r^2}-1+{{i\pi}\over 2}g_1.
\label{IV.13b}
\end{equation}
Furthermore, a particular solution to Eq. (\ref{IV.11}) is
\begin{equation}
g_0=2\left\{4\ln{(1+r^2)}-5+r^2+{{r^2-1}\over{r^2+1}}
\left[2{\rm dilog}(1+r^2)+\ln{r^2}\right]\right\},
\label{IV.14}
\end{equation}
where
\begin{equation}
{\rm dilog}(x)=\int_{1}^{x} {\ln{t}\over{1-t}}dt
\label{IV.15}
\end{equation}
is the dilogaritm function (\cite{AS} p. 1004).\par
Now coming back to the original function $\phi$ (see (\ref{IV.10})), we are 
led to the following instanton-like approximate solution to Eq. (\ref{IV.1}) 
(see (\ref{IV.5}), (\ref{IV.8})):
\begin{eqnarray}
W&=&\ln{4\over{(1+r^2)^2}}+\epsilon\left\{ c_1{{r^2-1}\over{r^2+1}}+
c_2\left({1\over 2}{{r^2-1}\over{r^2+1}}\ln{r^2}-1\right)+\right.\nonumber\\ 
&+&\left. 8\ln{(1+r^2)}-10+2{{r^2-1}\over{r^2+1}}\left[2{\rm dilog}(1+r^2)
+\ln{r^2}\right]-1\right\},
\label{IV.16}
\end{eqnarray}
$c_1$, $c_2$ being arbitrary constants. This formula is meaningful for 
small $\epsilon$ (large $\gamma$).\par
In order to confer an instanton character to the solution (\ref{IV.16}), we 
need to choose $c_2=-4$. Thus, the function $W$ becomes singularity free, 
precisely:
\begin{equation}
W=\ln{4\over{(1+r^2)^2}}+\epsilon\left\{ {{r^2-1}\over{r^2+1}}
\left[ c_1+4{\rm dilog}(1+r^2)\right]+8\ln (1+r^2)-7\right\}.
\label{IV.17}
\end{equation}
For small values of $\epsilon$, the function (\ref{IV.17}) can be used as a 
good approximation of the regular solution $W_I(r)$ of Eq. (\ref{III.2.30}) 
expressed in terms of $r$. In Appendix C we report two Tables, Ia and Ib, 
containing values of $W(r)$ and $W_I(r)$ for some $r$ with $\gamma$ fixed. 
The initial conditions have been chosen in such a way that $c_1=-7$ and 
$c_2=-4$. From the Tables one argues that $W(r)$ and $W_I(r)$ are quite 
closed already for $\gamma=6$ (i.e. $\epsilon=0.25$). The quantity 
$\vert W(r)-W_I(r)\vert$ tends to zero as $\gamma$ increases.

\section{Extension of Eq. (\ref{I.2}) to $(n+1)$-dimensions}
\setcounter{equation}{0}

Let us consider the generalized version
\begin{equation}
\left(\partial^2_{x_1}+\cdots +\partial^2_{x_n}\right)\;u=
k\left[\left(1+{u\over \gamma}\right)^{\gamma-1}\right]_{zz}
\label{V.1}
\end{equation}
of Eq. (\ref{I.2}), where $u=u(x_1,\ldots,x_n,\,z)$, with 
$(x_1,\ldots,x_n)\in {\cal{R}}^n$.\par
Assuming that $u=u(r,\,z)$, where $r=\sqrt{x_1^2+\cdots+x_n^2}$, Eq. 
(\ref{V.1}) can be written as
\begin{equation}
u_{rr}+{{n-1}\over r}u_r=k\left[\left(1+{u\over\gamma}\right)^
{\gamma-1}\right]_{zz}.
\label{V.2}
\end{equation}
The application of the symmetry approach to Eq. (\ref{V.2}) provides the 
symmetry generator
\begin{equation}
G=k_1r\partial_r+\left[\left(k_1+{{\gamma-2}\over 2}k_2\right)+k_3
\right]\partial_z+k_2(u+\gamma)\partial_u,
\label{V.3}
\end{equation}
with $k_1$, $k_2$, $k_3$ arbitrary constants.\par
From (\ref{V.3}) we get the operators
\begin{eqnarray}
G_1 &=& r\partial_r+z\partial_z,\label{V.4a}\\
G_2 &=& {{\gamma-2}\over 2}z\partial_z+(\gamma+u)\partial_u,\label{V.4b}\\
G_3 &=& \partial_z,\label{V.4c}
\end{eqnarray}
which satisfy the commutation relations
\begin{equation}
[G_1,G_2]=0,\;\;\;\;[G_1,G_3]=-G_3,\;\;\;\;[G_2,G_3]=-{{\gamma-2}\over 2}G_3.
\label{V.5}
\end{equation}
The symmetry operator $G_2$ is of particular interest. In fact, it gives 
rise to the invariants $r'=r$, $W'(r')=W(r)$, with
\begin{equation}
W(r)=z^{-{2\over {\gamma-2}}}(\gamma+u),
\label{V.6}
\end{equation}
which leads to the ordinary differential equation
\begin{equation}
W_{rr}+{{n-1}\over r}W_r={{2(\gamma-1)k}\over {{\gamma}^{\gamma-2}
(\gamma-2)^2}}W^{\gamma-1}.
\label{V.7}
\end{equation}
Equation (\ref{V.7}) becomes
\begin{equation}
\psi_{rr}+{{n-1}\over r}\psi_r=g\psi^{\gamma-1},
\label{V.8}
\end{equation}
with $g={{2(\gamma-1)k}\over {(\gamma-2)^2}}$, by setting $W=\gamma\psi$. 
The link beetween $u$ and $\psi$ is given by (see (\ref{V.6}))
\begin{equation}
u=\gamma\left( z^{2\over {\gamma-2}}\psi-1\right).
\label{V.9}
\end{equation}\par
Now we look for a transformation of the type
\begin{equation}
r'={1\over r},\;\;\;\psi=f(r'){\widetilde{\psi}}(r'),
\label{V.10}
\end{equation}
which leaves Equation (\ref{V.8}) invariant, in the sense that
\begin{equation}
{\widetilde{\psi'}}_{r'r'}+{{n-1}\over {r'}}{\widetilde{\psi'}}=
g{\widetilde{\psi}}^{\gamma-1}.
\label{V.11}
\end{equation}
A straightforward calculation shows that this can be accomplished if
\begin{equation}
f(r')={r'}^{n-2},\;\;\;\;\;\gamma={{2n}\over {n-2}}.
\label{V.12}
\end{equation}
In this case Eq. (\ref{V.8}) takes the form
\begin{equation}
\psi_{rr}+{{n-1}\over r}\psi_r=g_0{\psi}^{{n+2}\over {n-2}},
\label{V.13}
\end{equation}
with
\begin{equation}
g_0={k\over 8}(n^2-4)\;\;\;\;\;\;\;(n\neq 2).
\label{V.14}
\end{equation}\par
Furthermore, it is easy to see that if $\psi(\tilde{r})$ is a solution of 
Eq. (\ref{V.13}), i.e.
\begin{equation}
\psi_{{\tilde{r}}{\tilde{r}}}+{{n-1}\over {\tilde{r}}}\psi_{\tilde{r}}=
g_0\psi^{{n+2}\over{n-2}},
\label{V.15}
\end{equation}
then $\psi(r)=\lambda^{{n-2}\over 2}\psi(\tilde{r})$, where 
$\tilde{r}=\lambda r$ and $\lambda$ is a real parameter, is also a 
solution.\par
To summarize, we have\par\noindent
{\it{Proposition}}. Equation (\ref{V.13}) is invariant under the conformal 
transformation
\begin{equation}
r'={1\over {\lambda r}}\;,\;\;\;
\psi(r)={r'}^{n-2}\lambda^{{n-2}\over 2}{\widetilde{\psi}}(r').
\label{V.16}
\end{equation}\par
The positive (spherically symmetric) regular solution to Eq. (\ref{V.13}) 
is given by
\begin{equation}
\psi(r)={a\over{(1+r^2)^{{n-2}\over 2}}},
\label{V.17}
\end{equation}
with $a=\left( -{{8kn}\over {n+2}}\right)^{{n-2}\over 4}$, where $k$ and 
$n$ are to be chosen in such a way that $a>0$.\par
We notice that Eq. (\ref{V.13}) comprises remarkable special cases, which 
are listed in the following Table:\par\noindent\vskip 0.8cm
\begin{center}
{\bf{Table II}}
\end{center}
\begin{center}
\begin{tabular}{|c|c|c|l|c|}
\hline
Case   & $n$ & $\gamma$ & Equation (\ref{V.13}) & 
Regular solution $\psi(r)$ ($k=-1$)\\ \hline
a & 1  & -2  & $\psi_{rr}=-{{3}\over 8}k\psi^{-3}$ & 
$\left(-{{8k}\over 3}\right)^{-{1/4}}(1+r^2)^{1/2}$\\ \hline
b & 3  & 6   & $\psi_{rr}+{2\over r}\psi_{r}={5\over 8}k\psi^5$ &
$\left(-{24\over 5}k\right)^{1/4}(1+r^2)^{-1/2}$\\ \hline
c & 4  & 4   & $\psi_{rr}+{3\over r}\psi_{r}={3\over 2}k\psi^3$ &
$\left(-{16\over 3}k\right)^{1/2}(1+r^2)^{-1}$\\ \hline
d & 6  & 3   & $\psi_{rr}+{5\over r}\psi_{r}=4k\psi^2$ & 
$-6k(1+r^2)^{-2}$\\ \hline
e & 10 & 5/2 & $\psi_{rr}+{9\over r}\psi_{r}=12k\psi^{3/2}$ &
${{4\cdot 10^2}\over 9}(1+r^2)^{-4}$\\
\hline
\end{tabular}\par\vskip 0.8cm
\end{center}
The solutions of the original Equation (\ref{V.7}), related to cases 
a), ..., e), arise from
\begin{equation}
u={{2n}\over{n-2}}\left(z^{{n-2}\over 2}\psi-1\right)
\label{V.18}
\end{equation}
(see (\ref{V.9})).\par
It is noteworthy that Eq. (\ref{V.13}) appears in the context of 
Differential Geometry \cite{Y}, \cite{G}, in particular it is related to the 
socalled Yamabe problem, which consists essentially in establishing when a 
Riemannian metric can be changed by a conformal (lenght) factor to have 
constant scalar curvature \cite{Ya}.\par
Moreover, some special cases of Eq. (\ref{V.13}) play a basic role in 
certain branches of physics. Precisely, the equation of case b) of the 
Table II constitutes a static and spherically symmetric version of a nonlinear 
wave equation discussed by Rosen some years ago \cite{R}. Equation b), which 
is of the Emden type, finds application in Astrophysics.\par
Equation e) can be interpreted as an extended (elliptic) conformal 
invariant version of the equation
\begin{equation}
\Phi_{rr}+{2\over r}\Phi_r=C\Phi^{3/2}
\label{V.19}
\end{equation}
governing the Thomas--Fermi model of an atom, where $C>0$ is a certain 
constant and $\Phi(r)$ is the potential field originated by $Z-1$ 
electrons, acting on the $Z^{\rm th}$ one ([...], p. 125).\par
Equation a) corresponds to a potential whose nonlinear part, 
$1/(1-{u\over 2})^2$, mimics the inverse square potential appearing in the 
treatment of the scattering states in conformally invariant Quantum 
Mechanics \cite{BPS}. Finally, Eq. d) is associated with the Fermi--Pasta--Ulam 
potential.\par
To conclude the group analysis of Eq. (\ref{V.2}), we mention another 
nontrivial reduction, which can be written down starting from the generator 
$G_0=G_1+G_2+G_3$. It reads
\begin{eqnarray}
& X^2W_{XX}+(n-1)(W-XW_X)={{4(\gamma-1)}\over{{\gamma}^{\gamma-1}}}
{1\over{X^{\gamma-2}}}W^{\gamma-2}W_XX+\nonumber\\
& {{2(\gamma-1)(\gamma-2)}\over{{\gamma}^{\gamma-1}}}
{{W^{\gamma-3}W_X}\over{X^{\gamma-1}}}(2XW_X+W),
\label{V.20}
\end{eqnarray}
where $X={{(\gamma z+2)^{2\over{\gamma}}}\over r}$ and
$W=W(X)={{u+\gamma}\over{r}}$ are two invariants. Equation (\ref{V.20}) 
will not be discussed here.\par

\section{Relation to the Yang--Mills theory}
\setcounter{equation}{0}

Equation c) of Table II can be considered as a reduction form of the 
Yang--Mills equations (see later). This is a well--known result \cite{G}, 
\cite{tH}. What is new, is the link between special solutions of the 
Yang--Mills equations and special solutions of the field equation (\ref{V.1}) 
for $n=4$. To this aim, let us start from Eq. (\ref{V.8}), which can be 
written as
\begin{equation}
\Omega_{\tau\tau}+{{n-2}\over{\alpha-1}}\left( {{n-2}\over{n-2}}-\alpha
\right)\Omega_{\tau}-2{{n-2}\over{(\alpha-1)^2}}\left( \alpha-{n\over{n-2}}
\right)\Omega+\Omega^{\alpha}=0,
\label{VI.1}
\end{equation}
via the transformations
\begin{equation}
r=e^{-\tau},\;\;\;\Omega={1\over\mu}r^{2\over{\alpha-1}}\psi(r),
\label{VI.2}
\end{equation}
with $\mu=\left[ {{(\alpha-1)^2}\over{2\alpha}}\right]^{1\over{\alpha-1}}$, 
and $\alpha=\gamma-1$. We notice that Eq. (\ref{VI.1}) is the same as Eq. 
(1.6) contained in \cite{G}. Therefore, all the results achieved about this 
equation can be transferred to the field equation (\ref{V.1}) through the 
formula (\ref{V.9}), namely
\begin{equation}
u=(\alpha+1)\left[\left({{\alpha-1}\over{\sqrt{2\alpha}}}{z\over r}
\right)^{2\over{\alpha-1}}\Omega-1\right],
\label{VI.3}
\end{equation}
where $\Omega$ and $r$ are expressed by (\ref{VI.2}).\par
In general, Eq. (\ref{VI.1}) may be investigated using phase--space 
variables.\par
Here we restrict ourselves to the case $\alpha={{n+2}\over{n-2}}$, in 
correspondence of which Eq. (\ref{VI.1}) reads
\begin{equation}
\Omega_{\tau\tau}-\left({{n-2}\over{2}}\right)^2 \Omega+
\Omega^{{n+2}\over{n-2}}=0.
\label{VI.4}
\end{equation}
Equation (\ref{VI.4}) can be solved exactly. Indeed, we easily find
\begin{equation}
\Omega_{\tau}^2-\left({{n-2}\over 2}\right)^2\Omega^2+{{n-2}\over n}
\Omega^{{2n}\over{n-2}}=c,
\label{VI.5}
\end{equation}
where $c$ is a constant of integration.\par
By choosing for example $c=0$, a simple calculation gives
\begin{equation}
\Omega(r)=\left[\lambda{{\sqrt{n(n-2)}\cdot r}\over{r^2+\lambda^2}}
\right]^{{n-2}\over 2}
\label{VI.6}
\end{equation}
or, in the variable $\psi(r)$ (see (\ref{VI.2})):
\begin{equation}
\psi(r)=\left[4\lambda \sqrt{ {n\over{2(n+2)}}} {1\over{r^2+\lambda^2}}
\right]^{{n-2}\over 2}.
\label{VI.7}
\end{equation}
This function represents a regular solution to the equation
\begin{equation}
\psi_{rr}+{{n-1}\over r}\psi_r+{{n^2-4}\over 8}\psi^{{n+2}\over{n-2}}=0,
\label{VI.8}
\end{equation}
which comes from Eq. (\ref{V.8}) for $\alpha=\gamma-1={{n+2}\over {n-2}}$. 
On the other hand, replacing (\ref{VI.6}) into (\ref{VI.3}) provides the 
exact regular solution
\begin{equation}
u={{2n}\over{n-2}}\left\{ \left[ 4\lambda\sqrt{{n\over{2(n+2)}}}
{z\over{z^2+\lambda^2}}\right]^{{n-2}\over 2}-1\right\}
\label{VI.9}
\end{equation}
to the field equation (\ref{V.1}) $\left(\gamma={{2n}\over{n-2}}\right)$.\par
Equation (\ref{VI.4}) admits also singular solutions \cite{G}. The link 
between this kind of solutions and the corresponding ones of Eq. 
(\ref{V.1}) can be obtained via (\ref{VI.3}). At this stage, we point out 
that for $n=4$, a connection can be established between solutions of Eq. 
(\ref{V.1}) and solutions of the Euclidean Yang--Mills (YM) equations. To 
pursue this goal, let us introduce some preliminaries.\par
The pure YM equations (i.e., in absence of matter fields) are
\begin{equation}
\partial_\mu F_{\mu\nu}(x)+[A_\mu(x),\,F_{\mu\nu}(x)]=0
\label{VI.10}
\end{equation}
$(\mu\ \nu=0,\,1,\,2,\,3)$, where the summation convention is understood.\par
The fields $A_\mu(x)$ take values on the Lie algebra $\cal{G}$ of a compact 
semisimple Lie group $G$ (the gauge group). By choosing ${\rm{SU}}(2)$ as the 
gauge group, we can write
\begin{equation}
A_{\mu}(x)={1\over{2i}}A_{\mu}^a(x)\sigma^a
\label{VI.11}
\end{equation}
$(a=1,\,2,\,3)$, where $\sigma^a$ are the Pauli matrices and
$A_{\mu}^a=i{\rm{Tr}}\{ A_{\mu}(x)\sigma^a\}$.\par
On the othe hand, the tensor $F_{\mu\nu}$ associated with $A_{\mu}(x)$, 
defined by
\begin{equation}
F_{\mu\nu}(x)=\partial_{\mu}A_{\nu}-\partial_{\nu}A_{\mu}+[A_{\mu},\,A_{\nu}],
\label{VI.12}
\end{equation}
is expressed by
\begin{equation}
F_{\mu\nu}(x)={1\over{2i}}F_{\mu\nu}^a(x)\sigma^a,
\label{VI.13}
\end{equation}
with $F_{\mu\nu}^a(x)=i{\rm{Tr}}\{ F_{\mu\nu}(x)\sigma^a \}$.\par
Now, inserting the 't Hooft ansatz \cite{tH}
\begin{equation}
A_\mu^a=-\eta_{a\mu\nu}\partial_\nu\ln\psi(x)
\label{VI.14}
\end{equation}
into Eq. (\ref{VI.10}), where the tensor $\eta_{a\mu\nu}$will be specified 
later (see Appendix D), we obtain the equation
\begin{equation}
\nabla^2\psi+K\psi^3=0,
\label{VI.15}
\end{equation}
$K$ being a constant. By writing $\nabla^2$ in spherical coordinates and 
putting $K={2\over 3}$, Eq. (\ref{VI.15}) becomes just Eq. (\ref{VI.8}) for 
$n=4$. Then, by virtue of (\ref{VI.7}), (\ref{VI.2}), (\ref{VI.6}) and 
(\ref{VI.3}), from (\ref{VI.14}) we find the solution to the Yang--Mills 
equations (\ref{VI.12}):
\begin{equation}
A_\mu^a=-\eta_{a\mu\nu}\partial_\nu\ln\left[{1\over z}
\left(1+{u\over 4}\right)\right],
\label{VI.16}
\end{equation}
where $u$ is the regular solution
\begin{equation}
u=4\left({{4\lambda}\over{\sqrt{3}}}{z\over{r^2+\lambda^2}}-1\right)
\label{VI.17}
\end{equation}
to the nonlinear field equation (see (\ref{V.2}))
\begin{equation}
u_{rr}+{3\over r}u_r=-\left[\left(1+{u\over 4}\right)^3\right]_{zz},
\label{VI.18}
\end{equation}
with $r=\sqrt{x_1^2+x_2^2+x_3^2+x_4^2}$.\par
For completeness, in Appendix D a detailed proof of the correspondence 
between Eq. (\ref{VI.15}) and (\ref{VI.12}) is presented.

\section{Conclusion}
\setcounter{equation}{0}

Using a group--theoretical approach, we have investigated the nonlinear
field equation (\ref{I.2}), arising from a lattice of the binomial type as a 
continuous approximation, and its extension (\ref{V.1}) to 
$(n+1)$--dimensions. For $\gamma\rightarrow\infty$, both equations take the 
form of Toda field equations. We have determined the symmetry algebra 
(\ref{II.5a})--(\ref{II.5e}) admitted by Eq. (\ref{II.2}). This algebra is 
finite--dimensional, while that allowed by the Toda field equation 
(\ref{I.1}) is infinite--dimensional. This result constitutes a first 
discrimination between the two equations. Another important difference is 
that Eq. (\ref{I.1}) enjoys the conformal invariance property. On the 
contrary, this property is not valid for Eq. (\ref{I.2}), but it is regained 
through Eq. (\ref{V.13}) $(n\neq 2)$. In other words, the conformal 
invariance property holds again within the generalized equation (\ref{V.1}). 
This feature enables us to build up conformal versions of physically 
interesting models, listed in Table II. These conformal models afford both 
regular and singular solutions reflecting on the solutions of Eq. (\ref{V.1}) 
via the transformation (\ref{V.18}), which comes from the reduction procedure 
applied to Eq. (\ref{V.2}) (the spherically symmetric form of Eq. 
(\ref{V.1})). The introduction of the generalized equation (\ref{V.1}) leads 
to a scenario where several nonlinear field equations appearing in different
physical and differential geometrical contexts can be dealt with in a 
unifying manner. This is the spirit of the present work, which has essentially 
a speculative character.

\section*{Appendix A: explicit form of $pr^{(2)}V$}
\setcounter{equation}{0}
\renewcommand{\thesection}{\Alph{section}}
\addtocounter{section}{-6}

The second prolongation at the left--hand side of Eq. (\ref{II.2}) is given 
by 
\begin{eqnarray}
&pr^{(2)}V=pr^{(1)}V+\phi^{xx}{\partial\over{\partial u_{xx}}}+
 \phi^{xy}{\partial\over{\partial u_{xy}}}+
 \phi^{xz}{\partial\over{\partial u_{xz}}}+\nonumber\\
&\phi^{yy}{\partial\over{\partial u_{yy}}}+
 \phi^{yz}{\partial\over{\partial u_{yz}}}+
 \phi^{zz}{\partial\over{\partial u_{zz}}}\label{A.1}
\end{eqnarray}
where
\begin{equation}
pr^{(1)}V=V+\phi^x{\partial\over{\partial u_x}}+
\phi^y{\partial\over{\partial u_y}}+
\phi^z{\partial\over{\partial u_z}}
\label{A.2}
\end{equation}
\begin{eqnarray}
&\phi^x=D_x(\phi-\xi u_x-\eta u_y-\zeta u_z)+
 \xi u_{xx}+\eta u_{xy}+\zeta u_{xz}\label{A.3a}\\
&\phi^y=D_y(\phi-\xi u_x-\eta u_y-\zeta u_z)+
 \xi u_{xy}+\eta u_{yy}+\zeta u_{yz}\label{A.3b}\\
&\phi^z=D_z(\phi-\xi u_x-\eta u_y-\zeta u_z)+
 \xi u_{xz}+\eta u_{yz}+\zeta u_{zz}\label{A.3c}\\
&\phi^{xx}=D_x^2(\phi-\xi u_x-\eta u_y-\zeta u_z)+
 \xi u_{xxx}+\eta u_{xxy}+\zeta u_{xxz}\label{A.3d}\\
&\phi^{xy}=D_xD_y(\phi-\xi u_x-\eta u_y-\zeta u_z)+
 \xi u_{xxy}+\eta u_{xyy}+\zeta u_{xyz}\label{A.3e}\\
&\phi^{xz}=D_xD_z(\phi-\xi u_x-\eta u_y-\zeta u_z)+
 \xi u_{xxz}+\eta u_{xyz}+\zeta u_{xzz}\label{A.3f}\\
&\phi^{yy}=D_y^2(\phi-\xi u_x-\eta u_y-\zeta u_z)+
 \xi u_{xyy}+\eta u_{yyy}+\zeta u_{yyz}\label{A.3g}\\
&\phi^{yz}=D_yD_z(\phi-\xi u_x-\eta u_y-\zeta u_z)+
 \xi u_{xyz}+\eta u_{yyz}+\zeta u_{yzz}\label{A.3h}\\
&\phi^{zz}=D_z^2(\phi-\xi u_x-\eta u_y-\zeta u_z)+
 \xi u_{xzz}+\eta u_{yzz}+\zeta u_{zzz}\label{A.3i}
\end{eqnarray}
and $D_x$, $D_y$, $D_z$ denote the total derivative operators
\begin{equation}
D_x={\partial\over{\partial x}}+u_x{\partial\over{\partial u}},\;\;\;
D_y={\partial\over{\partial y}}+u_y{\partial\over{\partial u}},\;\;\;
D_z={\partial\over{\partial z}}+u_z{\partial\over{\partial u}}.
\end{equation}

\section*{Appendix B: derivation of Eq. (\ref{III.2.53})}
\setcounter{equation}{0}
\renewcommand{\thesection}{\Alph{section}}
\addtocounter{section}{+1}

Let us start from (\ref{III.2.46}), which can be written as
\begin{equation}
{{d\sigma}\over{\sqrt{1+4{\sigma}^3}}}=ig_3^{1\over 3}d\eta
\label{B1}
\end{equation}
via the substitution $\psi=-g_3^{1\over 3}\sigma$. Then
\begin{equation}
\int_\sigma^\infty{{d\sigma'}\over{\sqrt{1+4{\sigma'}^3}}}=
{1\over 3}\int_0^\infty{{X^{-2/3}dX}\over{\sqrt{1+4X}}}-
{1\over 3}\int_0^{\sigma^3}{{X^{-2/3}dX}\over{\sqrt{1+4X}}}=
ig_3^{1\over 6}(\eta-\eta_\infty),
\label{B2}
\end{equation}
where $X=\sigma^3$ and $\eta$ is a constant.\par
Now, we recall that (\cite{GR}, p. 285, formula 3):
\begin{equation}
\int_0^\infty{{X^{\mu-1}dX}\over{(1+bX)^\nu}}=b^{-\mu}B(\mu,\,\nu-\mu)
\label{B3}
\end{equation}
with $|{\rm arg}\,b|<\pi$, $Re\nu>Re\mu>0$, where $B(x,\,y)$ is the beta 
function (Euler's integral of the first kind) defined by (\cite{GR}, p. 
948)
\begin{equation}
B(x,\,y)=\int_0^1t^{x-1}(1-t)^{y-1}dt.
\label{B4}
\end{equation}
Taking account of (\ref{III.1.8}) and (\ref{B4}) with $b=4$, $\mu={1\over 3}$ 
and $\nu={1\over 2}$, Eq. (\ref{B2}) implies
\begin{equation}
\int_\infty^\sigma{{d\sigma'}\over{\sqrt{1+4{\sigma'}^3}}}=
-{{4^{-1/3}}\over 3}B\left({1\over 3},\,{1\over 6}\right)+
\sigma\,{_{2}F_{1}}\left( {1\over 2},\,{1\over 3};\,{4\over 3};\,
-4{\sigma}^3\right)=ig_3^{1\over 6}(\eta_\infty-\eta).
\label{B5}
\end{equation}
Since $\sigma=\wp (ig_3^{1\over 6}(\eta_\infty-\eta);\,0,\,-1)$ (see 
(\ref{III.2.50}) and (\ref{III.2.51})), we obtain 
$ig_3^{1\over 6}(\eta_\infty-\eta)={\wp}^{-1}(\sigma)$ and, therefore, 
formula (\ref{III.2.53}).\par
With the help of the property $B(x,\,y)={{\Gamma(x)\Gamma(y)}\over
{\Gamma(x+y)}}$, the quantity $B\left({1\over 3},\,{1\over 6}\right)$ can 
be calculated explicitly. We have
\begin{equation}
B\left({1\over 3},\,{1\over 6}\right)={{\Gamma\left({1\over 3}\right)
\Gamma\left({1\over 6}\right)}\over{\Gamma\left({1\over 2}\right)}}=
2^{2\over 3}{{\Gamma^2\left({1\over 3}\right)}\over
{\Gamma\left({2\over 3}\right)}}\simeq 8.413
\label{B6}
\end{equation}
where the duplication formula for the gamma function, $\Gamma(2z)=
{1\over {\sqrt{2\pi}}}2^{2z-{1\over 2}}$ $\Gamma(z)\Gamma\left(z+{1\over 2}
\right)$, has been applied to $\left(z={1\over 6}\right)$ (\cite{AS}, p. 255).

\section*{Appendix C: some values of $W(r)$ and $W_I(r)$ for fixed $\gamma$}
\setcounter{equation}{0}
\renewcommand{\thesection}{\Alph{section}}
\addtocounter{section}{+1}

\begin{center}
{\bf{Table Ia}}
\end{center}
\begin{center}
\begin{tabular}{|c||c|c||c|c||c|c|}
\hline
$\gamma$ & $W_I(0.1)$ & $W(0.1)$ & $W_I(0.3)$ & $W(0.3)$ & $W_I(0.5)$ 
  & $W(0.5)$\\ 
\hline
 4       & 1.100      & 0.763    & 0.597      & 0.406    & 0.163
  &0.069    \\ 
\hline
 6       & 1.148      & 1.048    & 0.729      & 0.678    & 0.369
  & 0.345    \\ 
\hline
 8       & 1.164      & 1.117    & 0.773      & 0.750    & 0.438
  & 0.427    \\ 
\hline
10       & 1.172      & 1.145    & 0.795      & 0.782    & 0.572
  & 0.466    \\ 
\hline
12       & 1.177      & 1.159    & 0.809      & 0.800    & 0.493
  & 0.489    \\ 
\hline
$\infty$ & 1.196      & 1.196    & 0.862      & 0.862    & 0.575
  & 0.575    \\ 
\hline
\end{tabular}\par\vskip 0.8cm
\end{center}

\begin{center}
{\bf{Table Ib}}
\end{center}
\begin{center}
\begin{tabular}{|c||c|c||c|c||c|c|}
\hline
$\gamma$ & $W_I(1)$ & $W(1)$ & $W_I(2)$ & $W(2)$ & $W_I(3)$ & $W(3)$\\ 
\hline
 4       & -0.727   & -0.727 & -2.041   & -2.438 & -3.030   & -5.840\\ 
\hline
 6       & -0.364   & -0.364 & -1.426   & -1.487 & -2.208   & -2.418\\ 
\hline
 8       & -0.242   & -0.242 & -1.221   & -1.245 & -1.934   & -2.012\\ 
\hline
10       & -0.182   & -0.182 & -1.118   & -1.131 & -1.797   & -1.837\\ 
\hline
12       & -0.145   & -0.145 & -1.057   & -1.065 & -1.715   & -1.740\\ 
\hline
$\infty$ & 0        & 0      & -0.811   & -0.811 & -1.386   & -1.386\\ 
\hline
\end{tabular}\par\vskip 0.8cm
\end{center}

\section*{Appendix D: solutions of Eq. (\ref{VI.10}) via solutions of Eq. 
(\ref{VI.15})}
\setcounter{equation}{0}
\renewcommand{\thesection}{\Alph{section}}
\addtocounter{section}{+1}

The tensor $\eta_{a\mu\nu}$ appearing at the right--hand side of (\ref{VI.14}) 
is defined as
\begin{equation}
\eta_{aij}=\epsilon_{aij},\;\;\;\eta_{a\nu 0}=-\eta_{a0\nu}=\delta_{a\nu},
\;\;\;\eta_{a00}=0,
\label{C.1}
\end{equation}
where $\epsilon_{aij}$ denotes the Ricci tensor. Other properties of the 
symbol $\eta$ are reported in \cite{tH}. It is also convenient to use the 
matrices
\begin{equation}
\sigma_{\mu\nu}={1\over 2}\eta_{a\mu\nu}\sigma^a
\label{C.2}
\end{equation}
(here $a,\,i,\,j,\,\ldots=1,\,2,\,3$, and $\mu,\,\nu,\,\lambda,\,\ldots=
0,\,1,\,2,\,3$). \par
In order to write $F_{\mu\nu}$ in terms of $\sigma_{\mu\nu}$, we shall 
evaluate the commutator $[\sigma_{\mu\nu},\,\sigma_{\rho\lambda}]$. Taking 
into account (\ref{C.1}) and (\ref{C.2}), we find
\begin{equation}
[\sigma_{\mu\nu},\,\sigma_{\rho\lambda}]={i\over 2}\eta_{a\mu\nu}
\eta_{b\rho\lambda}\epsilon_{abc}\sigma^c,
\label{C.3}
\end{equation}
where the commutation rule $[\sigma^a,\,\sigma^b]=2i\sigma^c$ has been 
employed. On the other hand, we have \cite{tH}
\begin{eqnarray}
&\eta_{a\mu\nu}\eta_{b\rho\lambda}\epsilon_{abc}=
 \epsilon_{cab}\eta_{a\mu\nu}\eta_{b\rho\lambda}=\nonumber\\
&\delta_{\mu\rho}\eta_{c\nu\lambda}-\delta_{\mu\lambda}\eta_{c\nu\rho}-
 \delta_{\nu\rho}\eta_{c\mu\lambda}+\delta_{\nu\lambda}\eta_{c\mu\rho}.
 \label{C.4}
\end{eqnarray}
Substituting (\ref{C.4}) into (\ref{C.3}) yields
\begin{equation}
[\sigma_{\mu\nu},\,\sigma_{\rho\lambda}]=
i(\delta_{\mu\rho}\sigma_{\nu\lambda}-\delta_{\mu\lambda}\sigma_{\nu\rho}-
\delta_{\nu\rho}\sigma_{\mu\lambda}+\delta_{\nu\lambda}\sigma_{\mu\rho}),
\label{C.5}
\end{equation}
with the help of (\ref{C.2}).\par
Now, the YM field (\ref{VI.11}) can be written as (see (\ref{VI.14}) and
(\ref{C.2}))
\begin{equation}
A_\mu=i\sigma_{\mu\nu}\partial_\nu \ln\psi.
\label{C.6}
\end{equation}
Hence, the tensor $F_{\mu\nu}$ (see (\ref{VI.12})) takes the form
\begin{eqnarray}
&F_{\mu\nu}=i\sigma_{\nu\rho}\partial_\mu(\partial_\rho\ln\psi)-
 i\sigma_{\mu\sigma}\partial_\nu(\partial_\rho\ln\psi)-
 i(\partial_\rho\ln\psi)(\partial_\lambda\ln\psi)
 [\sigma_{\mu\rho},\,\sigma_{\nu\lambda}]=\nonumber\\
& i(\Delta_{\mu\rho}-\Delta_{\mu}\Delta_{\rho})\sigma_{\nu\rho}-
  i(\Delta_{\nu\rho}-\Delta_{\nu}\Delta_{\rho})\sigma_{\mu\rho}-
  i(\Delta_\rho)^2\sigma_{\mu\nu},
\label{C.7}
\end{eqnarray}
where the shorthand notation
\begin{equation}
\Delta_\nu=\partial_\nu\ln\psi,\;\;\;\Delta_{\nu\rho}=\partial_\nu
\partial_\rho\ln\psi,\;\;\;\Delta_\nu^2=\partial_\nu^2\ln\psi,\;\;\;
(\Delta_\nu)^2=(\partial_\nu\ln\psi)^2,...
\label{C.8}
\end{equation}
is used.\par
Keeping in mind (\ref{C.7}) and (\ref{C.6}), the quantities 
$\partial_\mu F_{\mu\nu}$ and $[A_\mu,\,F_{\mu\nu}]$ (see Eq. 
(\ref{VI.12})) 
can be elaborated as follows:
\begin{eqnarray}
&\partial_\mu F_{\mu\nu}=i(\Delta_{\mu\mu\rho}-\Delta_\mu^2\Delta_\rho-
 \Delta_\mu\Delta_{\mu\rho})\sigma_{\nu\rho}\nonumber\\
&-i(\Delta_{\mu\nu\rho}-\Delta_{\mu\nu}\Delta_\rho-\Delta_\nu\Delta_{\mu\rho})
 \sigma_{\mu\rho}-2i\Delta_\rho\Delta_{\mu\rho}\sigma_{\mu\nu},
\label{C.9}
\end{eqnarray}
\begin{eqnarray}
&[A_\mu,\,F_{\mu\nu}]=i\Delta_\rho[\sigma_{\mu\rho},\,F_{\mu\nu}]\nonumber\\
&=-(\Delta_\rho\Delta_{\mu\lambda}-\Delta_\rho\Delta_\mu\Delta_\lambda)
 [\sigma_{\mu\rho},\,\sigma_{\nu\lambda}]\nonumber\\
&+(\Delta_\rho\Delta_{\nu\lambda}-\Delta_\rho\Delta_\nu\Delta_\lambda)
 [\sigma_{\mu\rho},\,\sigma_{\mu\lambda}]\nonumber\\
&+\Delta_\rho(\Delta_\lambda)^2[\sigma_{\mu\rho},\,\sigma_{\mu\nu}].
\label{C.10}
\end{eqnarray}
Then, putting (\ref{C.9}) and (\ref{C.10}) into the YM equations (\ref{VI.10}) 
and exploiting the commutation rule (\ref{C.5}), after some manipulations we 
obtain
\begin{eqnarray}
&\sigma_{\nu\rho}[\Delta_{\mu\mu\rho}-\Delta_\mu^2\Delta_\rho+
 2\Delta_\lambda\Delta_{\rho\lambda}-\Delta_\rho\Delta_\lambda^2-
 2\Delta_\rho(\Delta_\lambda)^2]\nonumber\\
&+\sigma_{\mu\rho}(-\Delta_{\mu\nu\rho}+2\Delta_\rho\Delta_\nu\Delta_\mu)=0.
\label{C.11}
\end{eqnarray}
Since the tensor multiplying $\sigma_{\mu\rho}$ is symmetric under the exchange 
$\mu\leftrightarrow \rho$, while $\sigma_{\mu\rho}$ is antisymmetric, Eq. 
(\ref{C.11}) becomes
\begin{equation}
\Delta_{\mu\mu\rho}+2\Delta_\mu\Delta_{\mu\rho}-2\Delta_\rho
[\Delta_\mu^2+(\Delta_\mu)^2]=0,
\label{C.12}
\end{equation}
which can be reduced to the form
\begin{equation}
\partial_\rho\left(\ln{{\partial_\mu^2\psi}\over \psi}-\ln\psi^2\right)=0.
\label{C.13}
\end{equation}
Equation (\ref{C.13}) tells us that the function in the bracket is 
independent from $x_\rho$ for any value of $\rho=0,\,1,\,2,\,3$. 
Consequently, Eq. (\ref{VI.15}) arises $(\partial_\mu^2=\nabla^2)$, where $K$ 
is a constant of integration.


\begin{thebibliography}{26}


\bibitem{M}   V. K. MAKHANKOV, "Soliton Phenomenology," Kluwer, Dordrecht 
              1990.

\bibitem{S}   See, for example, M. V. SAVELIEV, {\it{Teoret. Mat. Fiz.}} 
              {\bf{92}} (1992), 457, and references therein; M. PRZANOWSKI, 
              {\it{J. Math. Phys.}} {\bf{32}} (1991), 1004; C. LEBRUN, 
              {\it{J. Diff. Geom.}} {\bf{34}} (1991), 223.

\bibitem{FS}  D. B. FAIRLIE AND A. B. STRATCHAN, {\it{Physica D}} {\bf{90}} 
              (1996), 1.

\bibitem{FP}  J. D. FINLEY AND J. F. PLEBANSKI, {\it{J. Math. Phys.}} 
              {\bf{20}} (1979), 1939; C. P. BOYER AND J. D. FINLEY, 
              {\it{J. Math. Phys.}} {\bf{23}} (1982), 1126.

\bibitem{P}   Q--H. PARK, {\it{Phys. Lett. B}} {\bf{236}} (1990), 429.

\bibitem{K}   B. KONSTANT, {\it{Adv. Math.}} {\bf{34}} (1979), 195.

\bibitem{LLS} M. LEO, R. A. LEO AND G. SOLIANI, {\it{Phys. Lett. A}} 
              {\bf{60}} (1977), 283.

\bibitem{T}   M. TODA, "Theory of nonlinear lattices," in Solid--State 
              Science, vol. 20, Springer, Berlin 1981.

\bibitem{FPU} E. FERMI, J. PASTA AND S. ULAM, Los Alamos Report L. A. in
              "Collected papers of Enrico Fermi," edited by E. Segr\`e, 
              University of Chicago, 1965, vol. 2, 978.

\bibitem{O}   J. P. OLVER, "Application of Lie Group to Differential 
              Equations," Springer, New York 1986, Chap. 2.

\bibitem{CDF} MAPLE V Release 4; J. CARMINATI, J. S. DEVITT AND G. J. FEE, 
              {\it{J. Symbolic Comp.}} {\bf{14}} (1992), 103.

\bibitem{Alf} E. ALFINITO, G. SOLIANI AND L. SOLOMBRINO, {\it{Lett. Math. 
              Phys.}} {\bf{41}} (1997), 379.

\bibitem{F}   See, for example S. FL\"{U}GGE, "Practical Quantum Mechanics 
              II," Springer, Berlin 1971.

\bibitem{EH}  T. EGUCHI AND A. J. HANSON, {\it{Phys. Lett. B}} {\bf{74}} 
              (1978), 249.

\bibitem{Y}   S.--T. YAU, {\it{Survey on partial differential equations in 
              Differential Geometry}}, in "Seminar on Differential 
              Geometry," Ed. by S.--T. Yau, p. 3, Princeton University Press 
              1982.

\bibitem{G}   B. GIDAS, {\it{Symmetry and isolated singularities of 
              conformally flat metrics and of solutions of the Yang--Mills 
              equations}}, in "Seminar on Differential Geometry," Ed. by 
              S.--T. Yau, p. 243, Priceton University Press 1982.

\bibitem{tH}  G. 't HOOFT, {\it{Phys. Rev. D}} {\bf{14}} (1976), 3432.

\bibitem{W}   P. WINTERNITZ, {\it{Lie groups and solutions of nonlinear 
              partial differential equations}}, CRM--1841 1993.

\bibitem{GR}  I. S. GRADSHTEYN AND I. M. RYZHIK, "Tables of Integrals, 
              Series and Products," Academic Press, New York and London 
              1980.

\bibitem{Ka}  E. KAMKE, "Differentialgeichungen," Band 1, Chelsea Publ. 
              Comp., New York 1971.

\bibitem{BK}  G. W. BLUMAN AND S. KUMEI, "Symmetries and Differential 
              Equations," Springer, New York 1989, p. 84.

\bibitem{BPS} V. BARONE, V PENNA AND P. SODANO, {\it{Ann. Phys.}} 
              {\bf{225}} (1993), 212.

\bibitem{Ki}  J. KILLINGBECK, {\it{Phys. Lett. A}} {\bf{67}} (1978), 13.

\bibitem{AS}  M. ABRAMOWITZ AND I. A. STEGUN, "Handbook of Mathematical 
              Functions," Dover, New York 1972.

\bibitem{Ya}  H. YAMABE, {\it{Osaka Math. J.}} {\bf{12}} (1960), 21; R. 
              SCHOEN AND S.--T. YAU, "Lectures on Differential Geometry," 
              International Press, Boston 1994.

\bibitem{R}   G. ROSEN {\it{J. Math. Phys.}} {\bf{6}} (1965), 1269.


\end{thebibliography}
\end{document}